# Critical Current Densities and *n*-values of MgB$_2$ Strands over a Wide Range of Temperatures and Fields

G Z Li[1], Y Yang, M A Susner, M D Sumption, and E W Collings

Center for Superconducting and Magnetic Materials, Department of Materials Science and Engineering, the Ohio State University, Columbus, OH 43210, U.S.A.

E-mail: li.1423@osu.edu

**Abstract**

Transport measurements of critical current density, $J_{ct}$, in monocore powder-in-tube MgB$_2$ strands have been carried out at temperatures, $T$, of from 4.2 K to 40 K, and in transverse fields, $B$, of up to 14 T. Processing methods used were conventional continuous-tube-forming-filling (*CTFF*) and internal-magnesium-diffusion (*IMD*). Strands with several powder compositions were measured, including binary (undoped) MgB$_2$, 2% carbon doped MgB$_2$, and 3% carbon doped MgB$_2$. Magnetization loops (*M-B*) were also measured, and magnetic critical current density, $J_{cm}$, values extracted from them. The transport, $J_{ct}(B)$ and magnetic, $J_{cm}(B)$, critical current densities were compared. Also studied was the influence of doping on the resistively measured irreversibility field, $B_{irr}$ and upper critical field $B_{c2}$. Critical current densities, $J_{ct}$, and *n*-values were extracted from transport measurements and were found to be universally related (for all *B* and *T*) according to $n \propto J_{ct}^m$ in which $m = 0.52 \pm 0.11$. Likewise *n* was found to be related to *B* according to $n \propto B^{-p}$ with a *T*-dependent *p* in the range of about 0.08~0.21. Further analysis of the field (*B*) and temperature (*T*) dependencies of *n*-value resulted in an expression that enabled $n(B,T)$, for all *B* and *T*, to be estimated for a given strand based on the results of transport $J_{ct}(B)$ measurements made at one arbitrarily chosen temperature.

## LIST OF SYMBOLS and RELATIONSHIPS

**Symbols**

| | |
|---|---|
| $B$ | Magnetic field strength |
| $B_{irr}$ | Resistively determined irreversibility field |
| $B_{irr,trans}$ | Apparent irreversibility field for magnetically induced current paths which must always include a portion transverse to the strand axis |
| $B_{c2}$ | Upper critical field |
| $B_0(t)$ | Temperature-dependent field-normalization quotient in the exponential form of $J_c(B,t)$ – see below |
| $B_{00}$ | Zero-temperature value of $B_0(t)$ |
| *CTFF* | Continuous tube forming/filling (process for strand fabrication) |
| $d_0$ | Inner diameter of the Nb chemical barrier (both *CTFF* and *IMD* strands) |
| $d_i$ | Inner diameter of the annular MgB$_2$ reaction layer (*IMD* strands) |
| $\Delta M$ | Full height of the magnetization loop normalized to the actual SC volume |
| $F_p$ | Bulk pinning force density |
| $F_{p,max}$ | Maximum value of $F_p$ vs $B$ or $B/B_{irr}$ |
| *IMD* | Internal magnesium diffusion (process for strand fabrication) |
| $J_c$ | Abbreviation for the words "critical current density" |
| $J_{ct}$ | Transport-measured critical current density |
| $J_{cm}$ | Magnetically measured critical current density |
| $J_{cte}$ | Transport critical current density found by normalized $I_{ct}$ to the whole area inside the Nb chemical barrier |
| $J_{c2}$ | Critical current density transverse to the strand axis (magnetically induced) |
| $J_c(B,t)$ | Field- and temperature dependent $J_c$ |
| $J_{ct}(B,t)$ | Transport-measured value of field- and temperature dependent $J_c$ |
| $J_{c0}(t)$ | Fitted zero-field temperature-dependent $J_c$ |
| $J_{ct0}(t)$ | Fitted zero-field temperature-dependent $J_c$ based on transport measured field and temperature dependent $J_c$ equation |
| $J_{c00}$ | Fitted value of $J_c$ at zero field and zero temperature |
| $J_{ct00}$ | Fitted zero-field and zero-temperature $J_c$ value based on transport measured zero-field $J_c$ equation |
| $T$ | Temperature |
| $T_c$ | Transition (critical) temperature |
| $t$ | Reduced temperature, $T/T_c$ |
| $n$ | Index of the electric-field *vs* current-density (*E-J*) curve in the vicinity of $J_{ct}$ expressed in the form, $E/E_c = (J/J_{ct})^n$ |
| $n_0$ | $n(B)$ extrapolated to zero-*B* |
| $m$ | Index of the empirical relationship $n \propto J_c^m$ |
| $N$ | Empirical coefficient in $n = N J_c^m$ |
| $p$ | Index of the derived relationship $n \propto B^{-p}$. |

**Relationships**

| | |
|---|---|
| Exponential form of the field- and temperature dependence of $J_c$ | $J_c(B,t) = J_{c0}(t)\exp(-B/B_0(t))$ |
| Fitted zero-field temperature dependence of $J_c$ | $J_{c0}(t) = J_{c00}(1-\alpha t^2)$ |
| Fitted temperature dependence of the constant $B_0$ | $B_0(t) = B_{00}(1-\beta t)$ |
| Fitting constants in the above | $\alpha$ and $\beta$ |
| Fitted expansion of the relationship $n = N J_c^m$ in the form, of a full field- and temperature dependence of $n$ | $n(B,t) = N[J_{c00}(1-1.8t^2)\exp\{-B/B_{00}(1-1.2t)\}]^m$ |

# 1. Introduction

$MgB_2$ with its critical temperature, $T_c$, of about 40 K, and hence ability to operate at temperatures beyond the range of the low temperature superconductors NbTi and $Nb_3Sn$, is finding more and more commercial uses in devices such as magnetic resonance imagers, fault current limiters, motors and generators [1]. Over the years, substantial efforts have been made to increase $MgB_2$'s critical current density, $J_c$, at 4.2 K and elevated temperatures and in magnetic field up to 16 T by introducing dopants (especially those based on carbon [2]) and by performing pre-reaction cold high pressure densification [3]. Attention has also been paid to strand fabrication techniques, particularly Hyper Tech Research's (HTR) "continuous tube filling and forming" (*CTFF)* process in which the precursor powder mixture is continuously dispensed onto a strip of metal prior to its being formed into a tube [4], and the "internal Mg diffusion" (*IMD*) process, which starts off with a Mg rod imbedded axially in a B-filled tube and continues with wire forming and final heat treatment, HT [5]. Compared to the *CTFF* and conventional powder-in-tube (*PIT*) processes which yield a system of randomly connected $MgB_2$ fibers [6] *IMD* produces a dense $MgB_2$ layer structure with better longitudinal connectivity [7].

Interest in $MgB_2$ has led to numerous studies focusing on property improvement, in some cases in terms of critical fields via doping, and more recently in terms of $J_c$. In the latter area, quantification of strand improvement has frequently been based on the results of transport measurements, but mostly performed at 4.2 K for convenience. Other authors have gauged property improvement in terms of magnetization-derived $J_c$ results, sometimes performed over a range of temperatures. Thus in general the characterization of $MgB_2$'s electromagnetic properties and their improvement has tended to be rather narrowly focused: in some cases on critical field improvement, others only in terms of either magnetic $J_{cm}$ or transport based $J_{ct}$. Those examining transport properties are usually content to restrict themselves to 4.2 K, and only selected publications include $n$-value data. However, to evaluate an $MgB_2$ strand for a particular application it is important to have a complete data set, one in which magnetic and transport data taken on the same strands are shown side by side, with the $J_c$s and $n$-values presented as a functions of temperature and field. For these data to be useful, the samples chosen should be representative of high quality $MgB_2$ strands, of the kind that could be considered for application. This paper is not intended to offer a prescription for $MgB_2$ strand improvement; its focus is on an evaluation of transport and magnetic properties sufficient to aid in the design of applications based on them. We have restricted ourselves to the $J_c$s and $n$-values of four carefully selected strands and only two heat-treatment temperatures, allowing us to present sufficiently complete data from these strands to be useful for evaluating $MgB_2$ for various applications. We also have chosen monofilamentary strands for study, convinced that the results would more closely represent the intrinsic properties of strands in general. We have chosen to consider for comparison; (i) an undoped strand, (ii) strands with two levels of C doping, and (iii) two important strand- processing methods -- *CTFF* and *IMD*.

The four strand types studied were: the *CTFF*-processed P0 (undoped), P2 (2% C-doped), P3 (3% C-doped) and the 2%-C-doped *IMD*-processed strand I2. Resistance versus temperature curves were obtained in fields of from 0 T to 14 T and, based on the "10%" and "90%" transitions between the normal and superconducting states, the results were deconvoluted to yield the temperature dependencies of the irreversibility field, $B_{irr}$, and upper critical field, $B_{c2}$. The strong influence of carbon doping was noted. Based on a series of magnetization loop (*M-B*) measurements out to 14 T at 4.2 K ~ 30 K four groups of $J_{cm}$ versus *B* curves were generated and compared with the results of $J_{ct}$ measurements. As expected from previous such comparisons [8] in the higher field ranges $J_{cm}(B)$ dropped much faster than $J_{ct}(B)$ as a result of $J_c$ anisotropy. As pointed out previously [8] and again below, the high field divergence of $J_{ct}$ and $J_{cm}$ is due to the fact that $J_{ct}$ is the directly measured strand-longitudinal $J_c$ while $J_{cm}$ is controlled by a weaker strand-transverse critical current density.

Finally we have looked at "*n*-value", a parameter critical for applications but often overlooked, especially at temperatures other than 4.2 K. Defined as the index of the electric-field current-density (*E-J*) curve in the vicinity of $J_c$ expressed in the form $E \propto J^n$, *n* is a measure of the sharpness of the superconducting transition. A high *n* value indicates a sharp transition from superconducting state towards the normal state. It can be affected by intrinsic as well as extrinsic properties [9] and hence has both fundamental and practical importance. For example as the *n*-value of a strand increases the useful persistent current of a coil wound from it becomes closer and closer to its transport-measured $J_c$ [10]. Although many authors have discussed the $J_c$ of $MgB_2$, only a relative few have focused attention on *n*-value. Those that have specifically considered *n*-value have measured its dependencies on (i) temperature, *T* [11], (ii) applied field strength, *B* [11][12][13], (iii) critical current or critical current density, $J_c$ [10][11][13][14]. Although no quantitative relationships between *n*-value and *T* and *B* were established, two of the research groups pointed out that $n \propto J_c^m$ with $m \approx 0.37$ and 0.4 [10] and 0.5-0.7 [11]. Below we consider this in more detail, for each of the strands studied in turn.

This paper is organized as follows: As stated above, undoped and C-doped monocore MgB$_2$ strands processed by each of two techniques (*CTFF* and *IMD*) were selected for study. and comparison based on magnetic and transport measurements. The first part of this article focuses on their $J_c$s and critical fields $B_{irr}$ and $B_{c2}$ and the second part on their *n*-values, all over a wide range of applied fields, *B*, and temperatures, *T*. Accordingly the first part of the work explores the effect of doping on $B_{irr}(T)$, $B_{c2}(T)$, $J_{ct}(B,T)$ and $J_{cm}(B,T)$, takes the opportunity to emphasize the divergences of $J_{ct}(B,T)$ and $J_{cm}(B,T)$ at high fields, and concludes by comparing the connectivities of the *CTFF*- and *IMD*-processed strands.

The second part of the paper focuses on the presentation and discussion of transport-measured *n*-values. Based on a detailed analysis of the present transport measurements we confirm that $J_{ct}$ and *n*-value are universally related (for all *B* and *T*) according to $n \propto J_{ct}^m$ in which for the present strands $m = 0.52 \pm 0.11$. Furthermore we show that *n* is related to *B* according to $n \propto B^{-p}$ with a *T*-dependent *p* in the range of about 0.1~0.2. Finally we present an analysis of the field- and temperature dependencies of $J_{ct}$ and *n*-value that results in an expression that enables *n(B,T)*, for all *B* and *T*, to be estimated for a given strand, based on the results of transport measurements made at one arbitrarily chosen temperature.

## 2. Experimental

*2.1. The Samples*

A series of three monofilamentary strands (designated P0, P2, and P3), typically 0.83 mm OD, with a Nb barrier and a monel outer sheath were manufactured by Hyper Tech Research, Inc. (HTR) using the *CTFF* process [4]. One other strand (I2) was made using the *IMD* technique [5] based on positioning a 1/8 inch (3.2 mm) OD Mg rod along the axis of a B-filled double tube of Nb and monel. All samples used plasma-synthesized B from Specialty Materials Inc in either pure form or doped with three concentrations of C (e.g. [15]). Samples of the B powder were sent to the LECO Corporation for C content determination. The strands were heat treated at HTR in a tube furnace. They were ramped to soak temperature in about 80 min and furnace cooled to room temperature in about 300 min. The strand specifications and conditions are listed in Table I.

*2.2. Transport Measurements*

The transport measurements of $J_{ct}$ were carried out in transverse magnetic fields, $B$, of up to 12 T in pool boiling liquid He at 4.2 K on samples 50 mm long with a gauge length of 5 mm. Measurements were also made in a variable-temperature insert at 10~35 K on 30 mm long samples using a gauge length of 4 mm and an electric field criterion of 1 µV/cm. In both cases the gauge-length/sample-length ratio was adequate to ensure complete current sharing into the $MgB_2$ layers. The insert, located in the bore of the superconducting magnet, was a Cu can 150 mm long and 52 mm in diameter. In it, the sample was mounted between a pair of massive Cu bus-bars 10 mm x10 mm in cross section and 100 mm in length separated by a G10 plate, all suspended from a pair of 30 mm long, 2 mm diameter, current leads. A heater was attached to the plate and sample temperature was measured by a Cernox temperature-sensitive resistor attached to one of the bus-bars near the sample position. Thermal equilibrium at each temperature of measurement was established between sample-block cooling (by exchange gas and conduction along the current leads) and resistive heating.

The $J_{ct}$s of the *CTFF* strands were the critical transport currents, $I_{ct}$, divided by the area within the Nb chemical barrier. That of the *IMD* strand was $I_{ct}$ divided by the area of the annular $MgB_2$ layer (generally known as the "layer $J_c$"). For comparison purposes a "non-barrier $J_{ct}$", viz. $J_{cte}$", can also be defined for the *IMD* strand, as the $I_{ct}$ normalized to the *whole* area inside the chemical barrier.

*2.3 Magnetic and Resistive Measurements*

A Quantum Design Model 6000 Physical Property Measuring System (PPMS) was used to characterize the magnetic and resistive properties of the four strand samples. Magnetic $J_{cm}$s at 4.2~30K were extracted from magnetization versus applied field (*M-B*) loops measured on samples about 5 mm long at ramp rates of 13 mT/s in transverse fields of up to 14 T. Upper critical fields, $B_{c2}$, and irreversibility fields, $B_{irr}$, were derived from the results of voltage (hence resistance) measurements across gauge lengths of 2~4 mm using a 5 mA DC current again in transverse fields of up to 14 T and at temperatures from 4.2 K to 38 K.

## 3. Transport and Magnetic Critical Current Densities, $J_{ct}$ and $J_{cm}$

Transport measurements of $J_{ct}$ and magnetic measurements of $J_{cm}$ as functions of $B$ and $T$ were made as described above. As shown in Figure 1(a) the MgB$_2$ core of the *CTFF* strand is a solid cylinder (albeit with a longitudinal fibrous macrostructure [6][8]), whereas the superconducting core of the *IMD* strand is a hollow cylinder of MgB$_2$ [7], see Figure 1(b). Thus calculation of the $J_{cm}$s requires the usual Bean-model-based expression for a solid cylinder in a transverse magnetic field and a variant of it appropriate to hollow cylinders, the equations in SI units are:

| | | |
|---|---|---|
| For the superconducting core of *CTFF* monocore strand | $$J_{cm} = \frac{3\pi \Delta M}{4 d_0}$$ | (1) |
| For the MgB$_2$ reaction layer of *IMD* strand, the "layer $J_{cm}$" [13-15] | $$J_{cm} = \frac{3\pi \Delta M}{4 d_0 (1 - \frac{d_i^3}{d_0^3})}$$ | (2) |

Here $\Delta M$ is the full height of the magnetization loop normalized to the volume inside the Nb chemical barrier, $d_0$ is the inside diameter of the barrier (*CTFF* and *IMD* strands), and $d_i$ is the inner diameter of the MgB$_2$ reaction layer (*IMD* strand). The results are presented in Figure 2 and Table II.

### 3.1 Transport $J_{ct}$ and Connectivity

Figure 2 (a)-(d) displays the magnetic and transport derived critical current densities of all four samples. Considering first the transport results only, the data are limited at lower fields by either strand instabilities (all strands in this study are monofilaments) or the current limitation of the probe (220 A). Looking to the form of the field dependence, it is generally accepted by now that $MgB_2$ is a normal-surface (grain-boundary) pinner whose $J_c(B)$ follows the well known Kramer-Dew-Hughes relationship [19]. Nevertheless over a broad intermediate field range, and to a first approximation, $\log(J_{ct})$ decreases linearly with field. So for this reason, and for analytical convenience (see below) we have chosen to fit the data of Figure 2 (at all fields, $B$, and temperatures $t \equiv T/T_c$) to

$$J_c(B,t) = J_{c0}(t)\exp(-B/B_0(t)) \qquad (3)$$

Where $J_{c0}(t)$ is the zero-field temperature-dependent $J_c$, and $B_0(t)$ is a temperature-dependent field-normalization quotient. Then for later use when analyzing the dependence of $n$-value on $J_c$, $B$, and $T$ we went on to fit the temperature dependencies of $J_{c0}(t)$ and $B_0(t)$ leading to

$$J_{c0}(t) = J_{c00}(1-\alpha t^2) \qquad (4a)$$

$$B_0(t) = B_{00}(1-\beta t) \qquad (4b)$$

where $J_{c00}$ represents (0T,0K) critical current density and $B_{00}$ is the 0 K value of $B_0$, and in which the fitting constants from Figure 2 were found to be $\alpha = 1.3$ and $\beta = 1.0$. (Subsequently, in order to better fit the $n(B,T)$ data of Figure 6, adjustments to $\alpha = 1.8$ and $\beta = 1.2$ were required). Also for later use values of $J_{c0}$, and $B_0$ at 10 K and the maximum bulk pinning force densities, $F_{p,max}$, at 4.2 K and 10 K have been extracted from the transport data of Figure 2 and are presented in Table II.

Previous analysis of a *CTFF* strand of $MgB_2$ doped with 5 mol% of 30 nm SiC, which gave an $F_{p,max,10K}$ (i.e. $F_{p,max}$ at 10 K) value of 4.53 GN/m$^3$ [20] when compared with the resistively measured connectivity, *K,* of a pellet prepared from the same powder ($K$ = 6.41%) [21], indicated that if a 100%-connected C-doped $MgB_2$ were possible it would have have a $F_{p,max,10K}$ of about 70 $GN/m^3$. Taking the next step and referring to average data from Table II (strands P2 and P3) which provides $F_{p,max,4.2K}/ F_{p,max,10K}$ = 1.41 we conclude that such a perfectly connected strand, if it existed, would have an $F_{p,max,4.2K}$ = 99 $GN/m^3$ a value which is in reasonable accord with the $F_{p,max,4.2K}$ = 90 $GN/m^3$ estimated by Matsushita for fully connected well characterized bulk $MgB_2$ [22]. It is instructive to compare these estimates to corresponding results for the *IMD* strand, whose $MgB_2$ reaction layer is very dense, albeit with the possible presence of blocking layers.

We then can compare, for example, the 2%-C-doped *CTFF* and 2%-C-doped *IMD* strands in terms of both $J_{ct0}$ and $K$, in Table II. For the *CTFF* strand P2 the transport $J_c$ at 10 K, $J_{ct0,10K}$ = 8.1x10$^5$ A/cm$^2$, that for *IMD* strand I2 is tabulated as 59.6x10$^5$ A/cm$^2$. But when normalized to the *whole* area inside the chemical barrier (see Table 1) although the $J_{ct0,10K}$ of I2 reduces to 25.6x10$^5$ A/cm$^2$ it is still a factor of 3 greater than that of P2. In other words, on the basis of "non-barrier $J_{ct}$", viz. $J_{cte}$", the superior connectivity of the reaction layer provides the *IMD* strand with a greater $J_{cte}$ than that of the *CTFF*. But, most interesting from a scientific point of view, is the direct comparison of the *IMD* "layer $J_c$" with the *CTFF* core $J_c$ which indicates just how much of an increase in transport properties is available for very dense $MgB_2$ structures.

*3.2 Transport and Magnetic Critical Current Densities*

Particularly noticeable in Figure 2 is the premature drop-off of $J_{cm}$ with increasing field strength. As explained in detail in [8] this effect is a result of the difference between the $J_c$ along the strand, which is of course $J_{ct}(B)$, and that transverse to the strand axis, say $J_{c2}(B)$. The *CTFF* PIT strand is known to have a fibrous macrostructure stemming from the elongation of the starting Mg powder particles during wire drawing [6][8]. Consequently it is less well connected transversely than longitudinally. Furthermore magnetic measurements with the field directed along the *CTFF* strand axis have shown that the transverse currents are associated with a lower transverse irreversibility field, $B_{irr,trans}$. In a magnetic $J_{cm}$ measurement the loop current that supports the magnetization is controlled by the smaller of $J_{ct}(B)$ and $J_{c2}(B)$. Thus although at low fields $J_{cm}$ is equal to (or may turn out to be a bit higher than) $J_{ct}$, as the applied field tends towards $B_{irr,trans}$, $J_{cm}$ drops further and further below $J_{ct}$.

For the *IMD* strand the divergence happens at higher fields, but is still present. In this case the irregular wall thickness of the cylindrical $MgB_2$ reaction layer, Figure 1(b), may constrain $J_{c2}(B)$ at higher fields by transverse path area reductions which cause the current transfer length to be larger than the sample size, much the same as macrostructural limitations do in *CTFF* conductors.

**4. The Critical Fields, $B_{c2}$ and $B_{irr}$ and Critical Temperatures, $T_c$**

The critical fields, $B_{c2}$, and irreversibility fields, $B_{irr}$, were derived from the results of resistance (voltage) versus temperature, $T$, measurements carried out in perpendicular fields of up to 14 T. A typical $R \sim T$ curve is shown in Figure 3 for strand I2 in a 2 T applied field using a DC current of 5 mA. The point at which the resistance is 90% of its extrapolated normal value, $R_n$, is $(B_{c2},T)$. Likewise $(B_{irr},T)$ is the point corresponding to 10% $R_n$. The resulting $B_{c2}$ versus $T$ and $B_{irr}$ versus $T$ data are shown in Figures 4(a) and 4(b), respectively. The critical fields extrapolate to zero at the critical temperature $T_c$, which for strands P0, P2, P3, and I2 are, respectively 38.2 K, 35.4 K, 33.2 K, and 34.8 K.

It is well known by now (e.g. [21]) that atomic substitution into the B sublattice increases charge-carrier scattering which (i) reduces Cooper-pair coupling, lowers $T_c$, and hence tends to lower $B_{c2}$, (ii) increases normal-state resistivity which has the opposite effect on $B_{c2}$. With carbon doping the latter effect is dominant at lower temperatures while at higher temperatures the former effect dominates as $T$ approaches $T_c$.

The steady decrease in $T_c$ from 38.2 K to 33.2 K corresponds to an increase in the nominal starting C level from zero to 3 mol% and indicates that proportionate levels of C are substituting into the B sublattice. Furthermore, strands P2 and I2 with 2 mol% C have the same $T_c$s (35.4 K and 34.8 K). Carbon substitution, according to the above prescription, also explains the behaviors of the critical fields. At temperatures below 20 K the substitution by C produces a strong increase in the critical fields. But in the vicinity of 20-25 K the critical field curves cross over as effects of the lower $T_c$s of the C-doped strands begin to be felt.

## 5. Empirical Relationships between *n*-Value and Critical Current Density, Applied Field Strength, and Temperature

*5.1 Introduction to n-value and $J_c$*

Extracted from the transport *V-I* data (the source of Figure 2) *n* is the index of the electric-field current-density (*E-J*) curve in the vicinity of $J_{ct}$ expressed in the form, $E/E_c = (J/J_{ct})^n$, in which $E$ is electric field (the voltage drop across the sample's gauge length) and $J$ is current density. In these measurements, the criterion $E_c$ was set to 1 $\mu$Vcm$^{-1}$ and data from $2E_c$ to $20E_c$ were collected to obtain the *n*-values.

Several groups of authors have measured the dependencies of $n$-value on (i) temperature, $T$ [11], (ii) applied field strength, $B$ [11][12][13], and (iii) critical current or critical current density, $J_c$ [10][11][13][14]. Although no quantitative relationships between $n$-value and $T$ and $B$ were established, two of the research groups [10][11] did point out that $n \propto J_c^m$. Kim et al [10] measured a pair of 0.83 mm diameter *CTFF*-processed monofilamentary Nb/monel-sheathed MgB$_2$ strands HT at 650$^o$C/30 min, one undoped and the other C-doped using an addition of 10 wt% malic acid. Analysis of their data yielded $m$-indices of 0.37 (undoped) and 0.40 (doped). Based on measurements of HIPed and resistively heated bulk samples of undoped and doped MgB$_2$ and ex-situ-processed doped and undoped PIT strands Martínez et al [11] reported $m$-indices of 0.5 and 0.7. Implicit in the results of Martínez et al [13] and Kitaguchi et al [14] were $m$-indices of 0.72 [13] and 0.69 [14]. It seems that the $m$-index does not vary much (0.56 ± 0.16, based on [10][11][13][14] and 0.52 ± 0.11, based on the present work, Figure 5). On the other hand, hidden strand-to-strand differences show up in the prefactor $N$ of the equality $n = NJ_c^m$. Thus in the above-cited studies we find $N$ varying widely, from 0.02 [13] through 0.04 [11], 0.24 [11], to 0.28 [10] and 0.56 [10], and in the present studies from 0.01 to 0.50. It seems that the prefactor $N$ responds to strand type, processing conditions, and measurement temperature.

*5.2 Field Dependence of the n-value*

To further investigate the relationship between $n$-value and $J_c$ and hence its dependence on field (0~12 T) and temperature (4.2~30 K) we performed transport measurements of $J_{ct}(B,T)$ and the associated $n(B,T)$-values and plotted the results as in Figure 5(a) for strand P0 and Figure 5(b) for all the strands. The data, fitted to $n = NJ_c^m$ give for each strand unique values of $N$ and $m$ which are independent of temperature and applied field strength. The $n$-values in this work are for high quality monofilamentary strands, and although not necessarily "intrinsic" $n$ values (those associated with measurements of pinning potentials), they do represent what can be expected from transport measurements of $MgB_2$ strands in the absence of gross extrinsic limitations.

The results of transport property (combined $J_{ct}(B)$ and $n$-value) measurements can also be displayed in plots of $n$-value versus $B$. Several of the above authors have presented such plots [11][12][13] all of which indicate that $\log(n) \propto -pB$ with a temperature dependent slope, $-p$. In agreement with this trend are the results of the present transport property measurements as plotted in the format $\log(n)$ versus $B$, Figure 6.

The $B$-dependence of $n$ can be quantified in the following way: Starting with the empirical relationship $n = NJ_{ct}^m$ we next insert $J_{ct} = J_{ct0}.exp(-B/B_0)$ and find:

$$\log(n) = \log(N) + m.\log[J_{ct0}\exp(-B/B_0)] \qquad (5)$$

$$= \log(NJ_{ct0}^m) + 0.434m.\ln[\exp(-B/B_0)]$$

$$= \log(NJ_{ct0}^m) - 0.434mB/B_0 \qquad (6a)$$

$$= \log(n_0) - pB \qquad (6b)$$

such that, in a plot of $\log(n)$ versus $B$, $\log(n_0) \equiv \log(NJ_{ct0}^m)$ is the intercept and $-p \equiv -0.434m/B_0$ is the slope. Simply stated, just as $n \propto J_c^m$ so is $n \propto B^{-p}$.

Considering by way of example just the 10 K results, the experimentally determined values of the extrapolated intercept, $\log(n_0)$, and the slope, $-p$, are compared with the expectation from Equation (2) in Table III, to demonstrate self-consistency. In that table the experimental values of $J_{ct0}$ and $B_0$ are from linear fits to the 10 K data of Figure 2 and the $m$- and corresponding $N$-values are from data-fits to Figure 5.

*5.3 Development of an Expression for n-Value over a Wide Range of Temperatures, T, and Fields, B, based on a Set of Transport-$J_c(B)$ Measurements at a Single Temperature*

The expression for $n(B,T)$ or $n(B,t)$ to be developed is based on the exponential approximation for $J_{ct}(B)$ and the observation (Figure 5) that for a given strand at all temperatures $\log(n) \propto \log(J_{ct})$ and hence that $n = NJ_{ct}^m$. Then since $N$ and $m$ are temperature independent the temperature dependence of $n(B,t)$ can be assigned to that of $J_{ct}(B,t)$, in other words :

$$n(B,t) = N[J_{ct00}(1-1.8t^2)\exp\{-B/B_{00}(1-1.2t)\}]^m \qquad (7)$$

This leads to the following prescription for predicting the temperature dependence of $n(B,t)$:

At some convenient fixed temperature, designated $T_{meas}$ (e.g. 10 K), and over a wide range of fields, measurements are made of $J_{ct}(B)$ and the corresponding $n(B)$, after which

(i) $n(B)$ is fitted to $NJ_{ct}^m(B)$ and values of $N$ and $m$ determined,

(ii) $J_{ct}(B)$ is fitted to $J_{ct}(B) = J_{ct0}.\exp(-B/B_0)$ and values of $J_{ct0}(T_{meas})$ and $B_0(T_{meas})$ determined,

(iii) values of $J_{ct00}$ and $B_{00}$ are determined following Equations 4(a) and 4(b), respectively.

The quantities $N$, $J_{ct00}$, $B_{00}$, and $m$ having been obtained in this way $n(B,t)$ can be determined by substitution into Equation (7).

By way of example we offer Figure 7. Using only 10 K data for samples P2 and P3 we use Equation (7) to predict $n(B)$ for five temperatures between 10 K and 20 K and compare the resulting curve with the experimentally measured $n(B)$ data.

## 6. Summary

As a useful starting point for the analysis of MgB$_2$ strands for application we have performed a comprehensive set of transport and magnetic measurements on strands representing state-of-the-art MgB$_2$ conductors over a wide range of temperatures and fields. Measured were three *CTFF*-processed strands, P0, P2, and P3, with nominal 0, 2 and 3 mol% C additions, and one *IMD*-processed strand, I2, with 2 mol% C addition. Transport and magnetic critical current densities ($J_{ct}$ and $J_{cm}$ respectively) were measured in perpendicular applied fields of up to 12 T and at temperatures in the range 4.2~40 K. The observed linear decrease of log($J_{ct}$) over a broad intermediate field range justified data fits to the expression $J_{ct}(B) = J_{ct0} \exp(-B/B_0)$. As the fields increased above the mid-range values rapid divergences of $J_{ct}(B)$ and $J_{cm}(B)$ were observed. The premature decreases in $J_{cm}(B)$ in the case of P0, P2, and P3, were attributed to weakness in the transverse $J_c$ caused by the strands' fibrous longitudinal macrostructure. A similar effect in the case of I2 was attributed to uneven wall thickness of the hollow cylindrical reaction layer. Extracted from the $J_{ct}(B)$ data of *CTFF* strands at an arbitrarily chosen 10 K values of the maximum bulk pinning force $F_{p,max,10K}$ were calculated to be in the 7-8 GN/m$^3$ range, and, based on estimated connectivity values would be expected to be roughly 100

GN/m$^3$ at 4.2 K if full connectivity were possible (compared to a previously estimated 90 GN/m$^3$ for a bulk sample of MgB$_2$). It is interesting to compare this to the actual measured values for the very dense *IMD* samples where the $F_{p,max}$ at 4.2 K is seen to reach 60 GN/m$^3$, relatively close to the expected value for fully connected strands with the given grain size. Even though the filament is hollow, as a consequence of the very high layer $J_{ct0}$ resulting from an estimated strong connectivity, the engineering $J_{ct0e}$ (that normalized to the whole area within the Nb barrier) of *IMD* strand I2 turned out to be a factor of 3 greater than the $J_{ct0}$ of *CTFF* strand P2.

The "90% and 10% transition points" of *R(T)* curves taken over a wide range of fields led to plots of $B_{c2}(T)$ and $B_{irr}(T)$ after which short extrapolations to zero field provided the critical temperatures, $T_c$. Carbon content and $T_c$ were directly related; $T_c$ decreased monotonically with increasing mol%C (suggesting C substitution into the B sublattice) and the $T_c$s of P2 and I2 were practically the same. Charge-carrier scattering by C-substitution tends to lower $T_c$ and increase normal-state resistivity. At lower temperatures the latter effect dominates and the critical fields increase with the addition of the C; the opposite occurs at higher temperatures as $T_c$ is lowered by C substitution.

Transport $J_{ct}(B,T)$ measurements also provided corresponding values of $n(B,T)$ which were subjected to detailed analysis. It turned out that for all fields and temperatures the *n*-data all condensed onto a single linear plot of log($n(B,T)$) versus $J_{ct}$ leading to $n \propto J_{ct}^m$, a relationship frequently noted by others. But by extending the analysis further we determined the constants $N$ and $m$ in the relationship $n = NJ_{ct}^m$. The function log($n(B,T)$) was also plotted versus $B$ for each temperature of measurement (4.2-30 K). and the resulting linearities described in terms of a proportionality $n \propto B^{-p}$ in which the index *p* is temperature dependent. Further analysis of the field (*B*) and temperature ($t = T/T_c$) dependencies of *n*-value resulted in the expression *n(B,t)* = $N[J_{ct00}(1-1.8t^2)exp\{-B/B_{00}(1-1.2t)\}]^m$ which enabled *n(B,T)* for all *B* and *T* to be estimated for a given strand based on the results of transport $J_{ct}(B)$ measurements made at one arbitrarily chosen temperature. We expect that this data, taken together with the transport data from the first part of the work, can be useful for choosing applications and operational regimes for $MgB_2$ conductors, and also useful for models of system performance.

**Acknowledgements**

The strands were manufactured by HyperTech Research Inc under the supervision of M. A. Rindfleisch, M. J. Tomsic. The research and strand manufacture were funded by the U.S. Department of Energy (DOE), Office of High Energy Physics, under Grant No. DE-FG02-95ER40900, DOE SBIR grants, and a grant from the Ohio Department of Development.

# LIST OF TABLES

Table I: Monofilamentary Strand Specifications and Conditions

| Name | Process | MgB$_2$ core* diam., μm | Dopant conc., mol%C** | Strand fill factor, % | Heat treatment at soak. |
|---|---|---|---|---|---|
| P0 | *CTFF* | 419 ($d_0$) | zero | 25.2 | 675°C /20min |
| P2 | *CTFF* | 325 ($d_0$) | 2.09 | 15.2 | 675°C/20min |
| P3 | *CTFF* | 302 ($d_0$) | 3.15 | 13.1 | 700°C /20min |
| I2 | *IMD* | 270 ($d_0$) <br> 204 ($d_i$) | 2.09 | 5.2 | 675°C /30min |

\* Mg:B atomic ratio, 1:2

\*\* Based on C analysis by the LECO Corporation and normalized to the molar weight of MgB$_2$. No assumptions are made here concerning the expected uptake of C into the B sublattice; see, however, [15].

**Table II: Transport Properties of the Strands**

| Strand | P0 | P2 | P3 | I2 |
|---|---|---|---|---|
| $J_{ct0,10K}$, $10^5$ A/cm$^2$ | 18.4 | 8.1 | 10.4 | 59.6 |
| $B_{0,10K}$, T | 1.07 | 1.86 | 2.30 | 1.90 |
| $F_{p,max,10K}$, GN/m$^3$ | 7.2 | 5.5 | 8.7 | 41.5 |
| $F_{p,max,4.2K}$, GN/m$^3$ | 5.6 | 8.4 | 11.3 | 60.9 |
| Estimated connectivity, $K$, % | 12 | 9 | 15 | 69 |

**Table III: Analysis of the 10 K Transport Property Results**

| Strand | P0 | P2 | P3 | I2 |
|---|---|---|---|---|
| Trans. temp. $T_c$, K | 38.2 | 35.4 | 33.2 | 34.8 |
| $J_{c0}$, $10^5$ A/cm$^2$ | 18.37 | 8.07 | 10.40 | 59.56 |
| $B_0$, T | 1.07 | 1.86 | 2.30 | 1.90 |
| $m_{10K}$ | 0.537 | 0.434 | 0.417 | 0.491 |
| $N_{10K}$ | 0.138 | 0.333 | 0.242 | 0.134 |
| $p_{10K} = 0.434 m/B_0$ | 0.208 | 0.115 | 0.084 | 0.145 |
| $p_{10K,measured}$, Figure 6 | 0.219 | 0.102 | 0.083 | 0.112 |
| $n_{0,10K, calculated} = N_{10K} J_{c0}^{m10K}$ | 316 | 121 | 78 | 286 |
| $n_{0,10K,measured}$, Figure 6 | 316 | 123 | 87 | 281 |

# LIST OF FIGURES

**Figure 1:** (a) A typical *CTFF* PIT monocore strand consisting of a $MgB_2$ core enclosed in a Nb chemical barrier and an outer monel sheath. (b) The *IMD*-processed Strand I2 consisting of an annulus of $MgB_2$ enclosed in a Nb chemical barrier and an outer monel sheath. The central core is the porous residue of the starting Mg rod.

**Figure 2:** Transport $J_{ct}$ and magnetic $J_{cm}$ versus $B$ in transverse applied fields at temperatures between 4.2 K and 30 K for (a) Strand P0 (undoped *CTFF*), (b) Strand P2 (2% C-doped *CTFF*), (c) Strand P3 (3% C-doped *CTFF*) and (d) Strand I2 (2% C-doped *IMD*). The $J_{ct}$ are represented by lines through data points; the $J_{cm}$ are represented by the "corresponding" full lines (arranged right-to-left in descending order of temperature).

**Figure 3:** By way of example, the resistance versus temperature plot for *IMD* strand P2 measured at 2 T indicating the 90% and 10% transition points.

**Figure 4:** Temperature dependence of: (a) the upper critical field, $B_{c2}$ and (b) the irreversibility field, $B_{irr}$.

**Figure 5:** (a) *n*-value versus transport $J_{ct}(B,T)$ for undoped strand P0 at different temperatures and (b) *n*-value versus transport $J_{ct}(B,T)$ for all strands at all fields (up to 12 T) and temperatures (4~30 K).

**Figure 6** *n* value versus $B$ in perpendicular applied fields at temperatures between 4.2 K and 40 K for: (a) undoped *CTFF* strand P0, (b) 2% carbon doped *CTFF* strand P2, (c) 3% carbon doped *CTFF* strand P3, and (d) 2% carbon doped *IMD* strand I2.

**Figure 7**: *n*-value versus $B$ at temperatures between 10 K and 20 K for *CTFF* strands P2 (a) and P3 (b) based on Equation (7) (full lines) compared to the experimental results (data points)

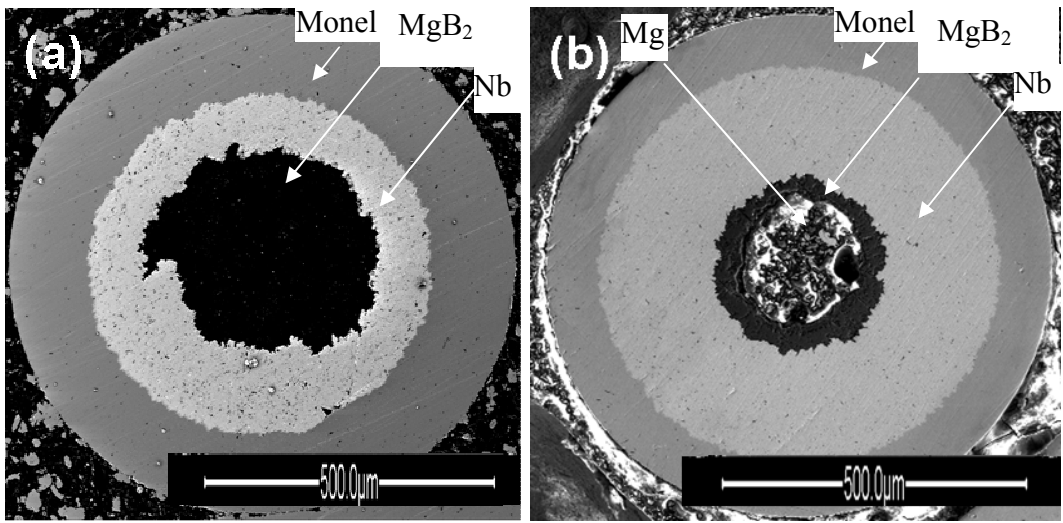

**Figure 1:** (a) A typical *CTFF* PIT monocore strand consisting of a $MgB_2$ core enclosed in a Nb chemical barrier and an outer monel sheath. (b) The *IMD*-processed Strand I2 consisting of an annulus of $MgB_2$ enclosed in a Nb chemical barrier and an outer monel sheath. The central core is the porous residue of the starting Mg rod.

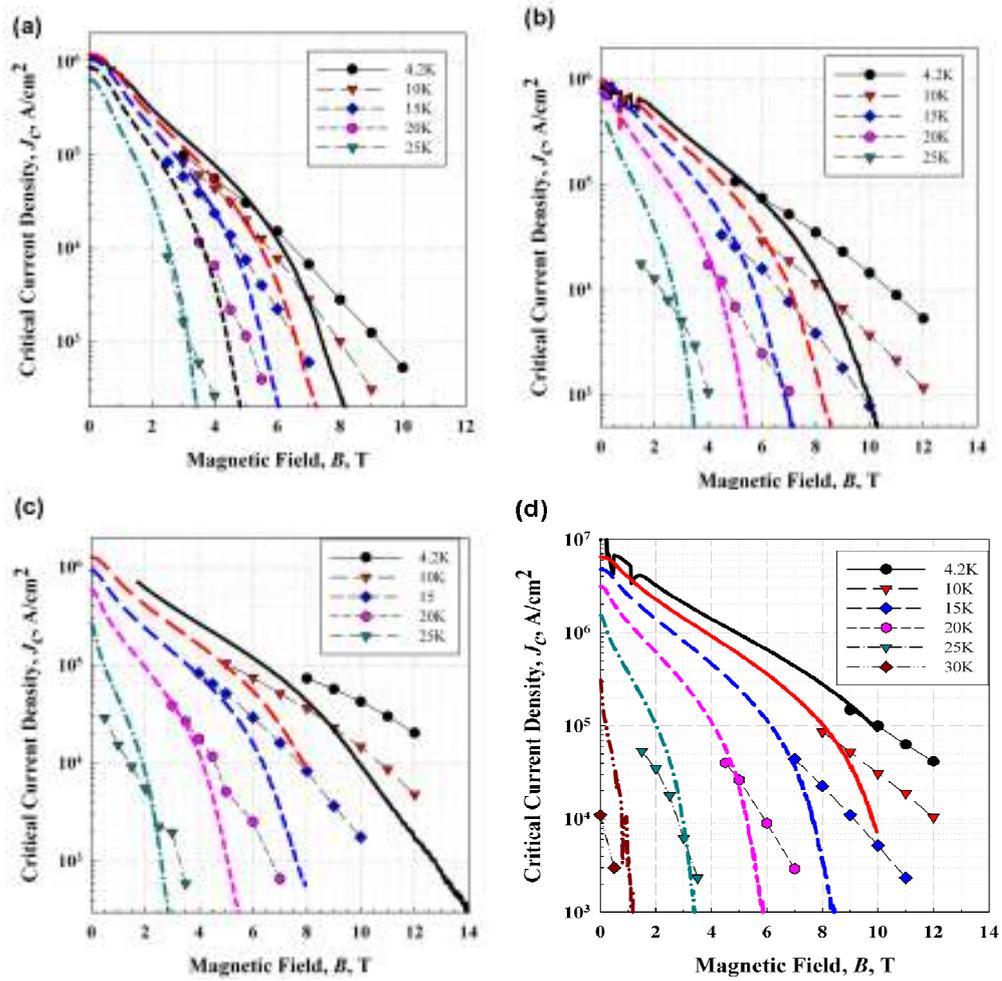

**Figure 2:** Transport $J_{ct}$ and magnetic $J_{cm}$ versus $B$ in transverse applied fields at temperatures between 4.2 K and 30 K for (a) Strand P0 (undoped *CTFF*), (b) Strand P2 (2% C-doped *CTFF*), (c) Strand P3 (3% C-doped *CTFF*) and (d) Strand I2 (2% C-doped *IMD*). The $J_{ct}$ are represented by lines through data points; the $J_{cm}$ are represented by the "corresponding" full lines (arranged right-to-left in descending order of temperature).

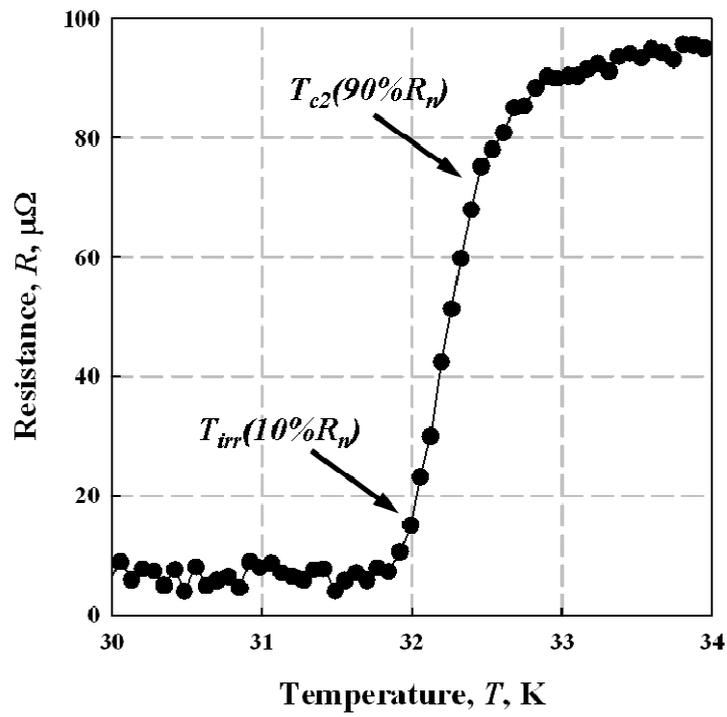

**Figure 3:** By way of example, the resistance versus temperature plot for *IMD* strand P2 measured at 2 T indicating the 90% and 10% transition points.

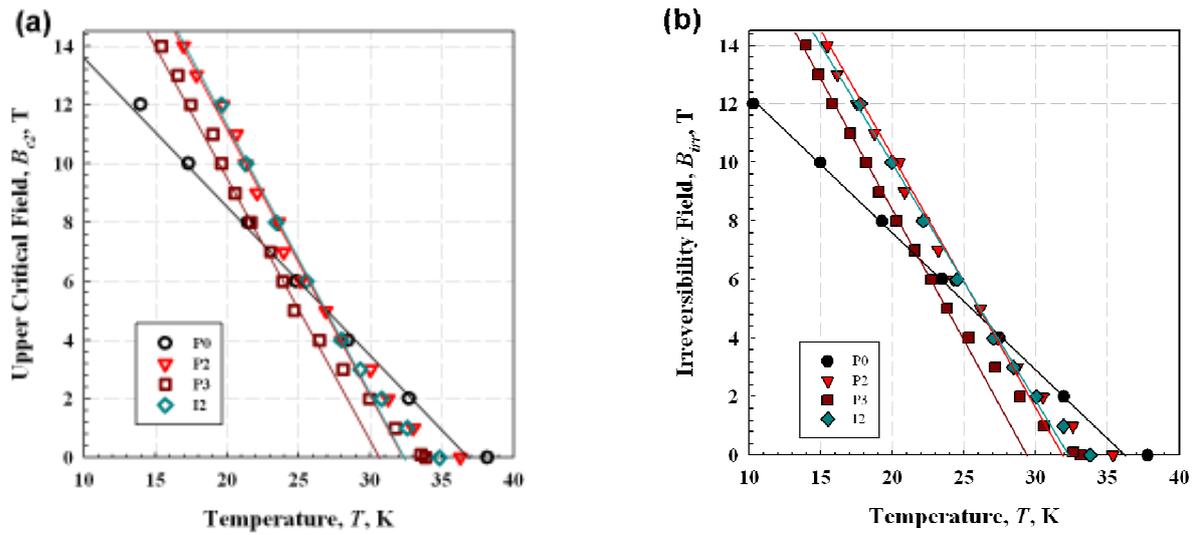

**Figure 4.** Temperature dependence of: (a) the upper critical field, $B_{c2}$ and (b) the irreversibility field, $B_{irr}$

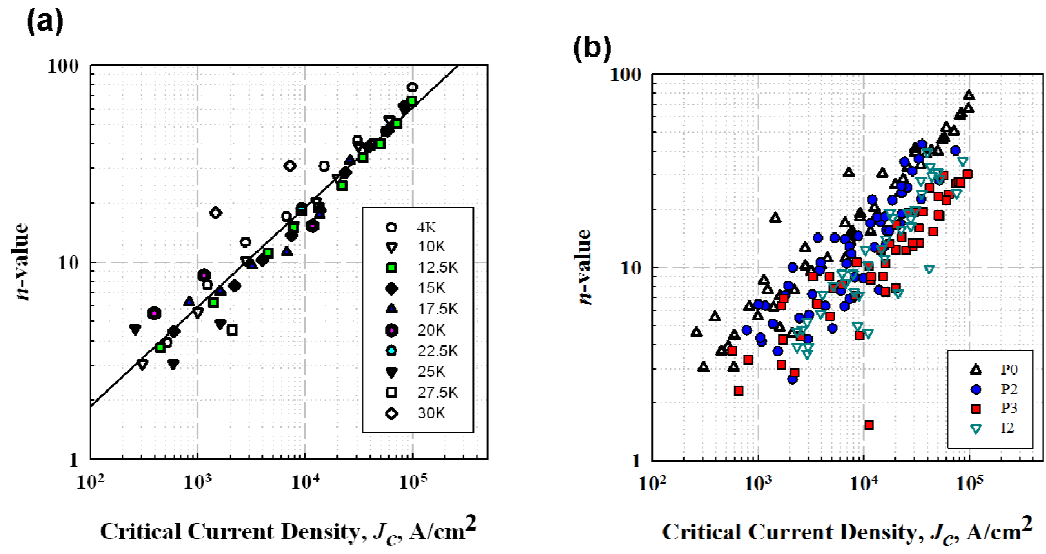

**Figure 5:** (a) *n*-value versus transport $J_{ct}(B,T)$ for undoped strand P0 for different temperatures and (b) *n*-value versus transport $J_{ct}(B,T)$ for all strands at all fields (up to 12 T) and temperatures (4~30 K)

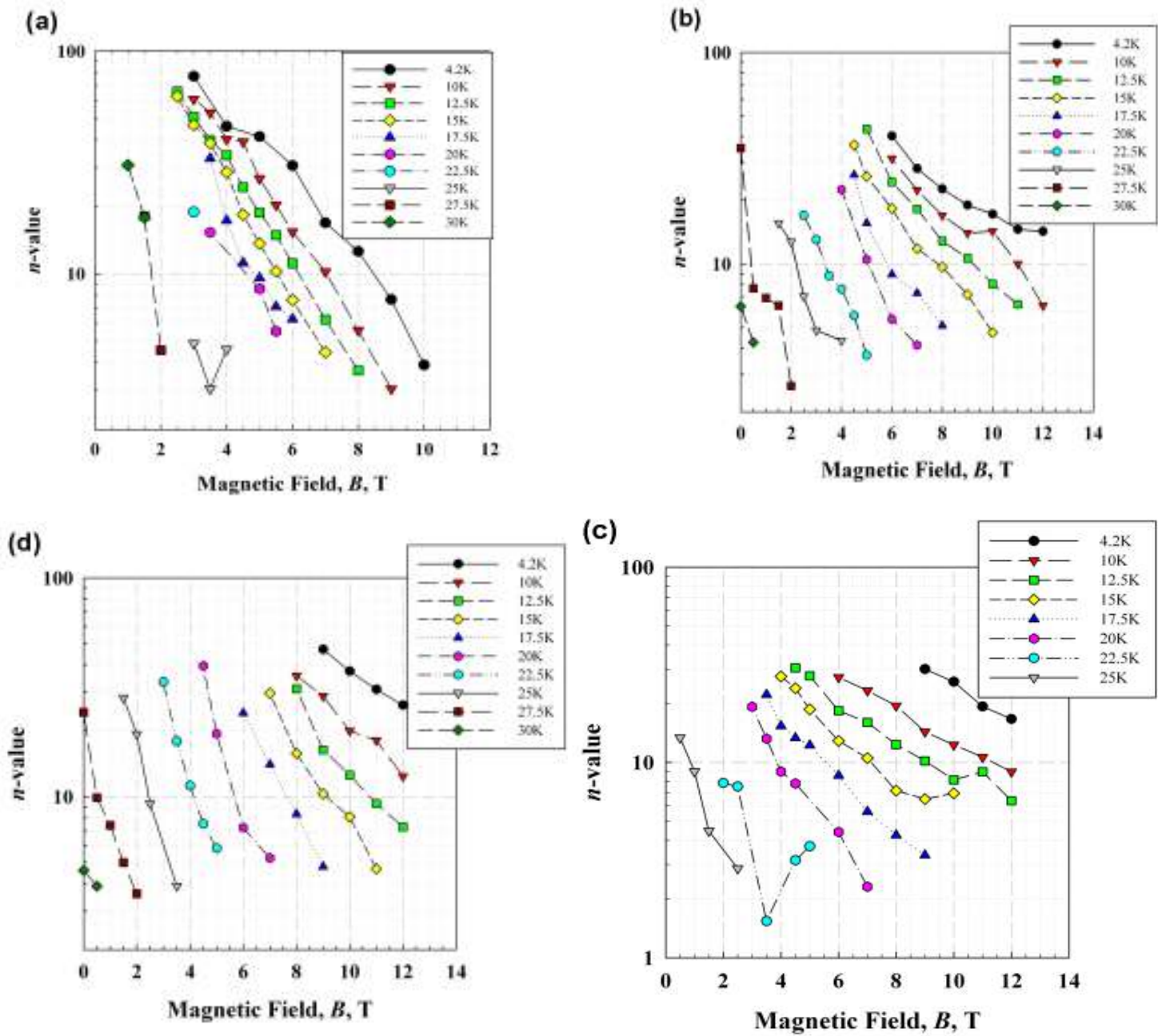

**Figure 6** *n* value versus *B* in perpendicular applied fields at temperatures between 4.2 K and 40 K for: (a) undoped *CTFF* strand P0, (b) 2% carbon doped *CTFF* strand P2, (c) 3% carbon doped *CTFF* strand P3, and (d) 2% carbon doped *IMD* strand I2.

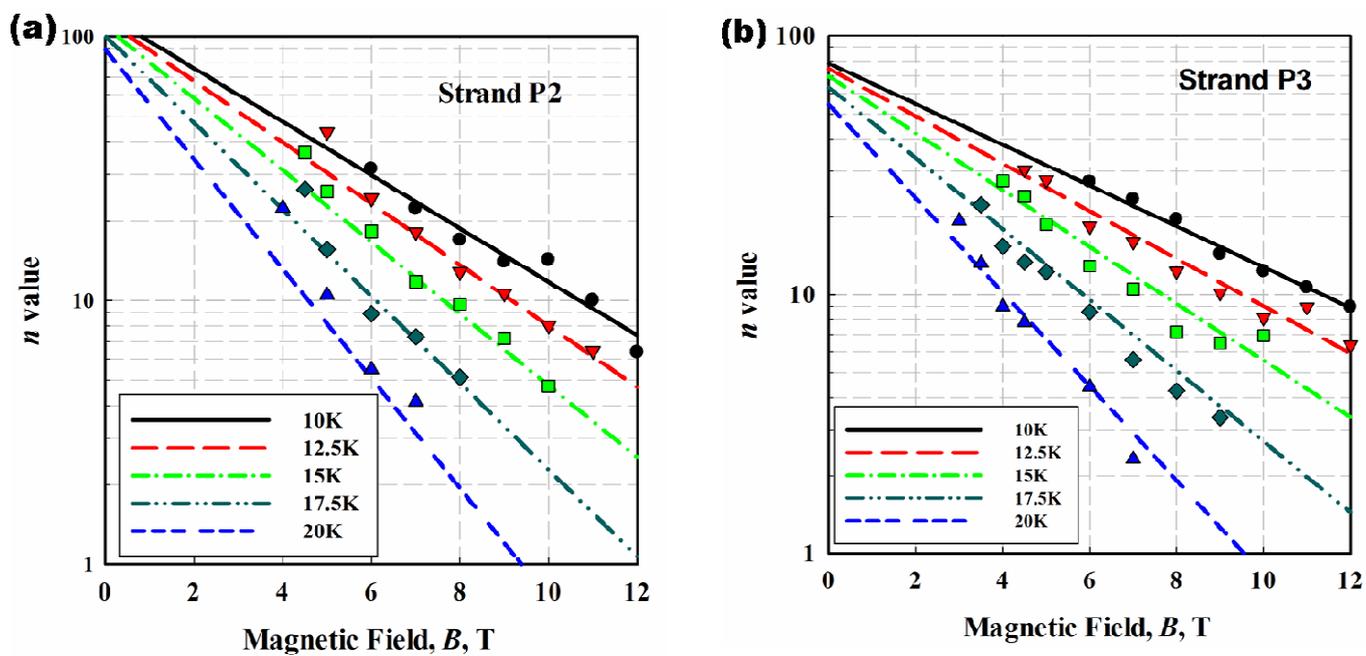

**Figure 7**: *n*-value versus *B* at temperatures between 10 K and 20 K for *CTFF* strands P2 (a) and P3 (b) based on Equation (7) (full lines) compared to the experimental results (data points)